\documentclass[aps,preprint]{revtex4}
\usepackage{latexsym}
\usepackage{graphicx}
\usepackage{epsfig,subfig}
\usepackage{epstopdf}
\usepackage[active]{srcltx}
\usepackage{amsmath}
\usepackage{amssymb}
\usepackage{hyperref}
\usepackage{cleveref}
\usepackage{color}
\usepackage{diagbox}
\usepackage{float}

\begin{document}

\title{The Polymer Bouncer}

\author{A. Martin-Ruiz}
\email{alberto.martin@nucleares.unam.mx}
\affiliation{Instituto de Ciencias Nucleares, Universidad Nacional Aut\'{o}noma de M\'{e}xico, 04510 M\'{e}xico, Distrito Federal, M\'{e}xico}

\author{A. Frank}
\affiliation{Instituto de Ciencias Nucleares, Universidad Nacional Aut\'{o}noma de M\'{e}xico, 04510 M\'{e}xico, Distrito Federal, M\'{e}xico}
\affiliation{Centro de Ciencias de la Complejidad, Universidad Nacional Autónoma de México, 04510 M\'{e}xico, Distrito Federal, M\'{e}xico}

\author{L. F. Urrutia}
\affiliation{Instituto de Ciencias Nucleares, Universidad Nacional Aut\'{o}noma de M\'{e}xico, 04510 M\'{e}xico, Distrito Federal, M\'{e}xico}

\begin{abstract}
Polymer Quantization (PQ) is a background independent quantization scheme that is deployed in Loop Quantum Gravity. This framework leads to a new short-distance (discretized) structure characterized by a fundamental length. In this paper we use PQ to analyze the problem of a particle bouncing on a perfectly reflecting surface under the influence of Earth's gravitational field, what we have called ``\textit{The Polymer Bouncer}". In this scenario, deviations from the usual quantum effects are induced by the spatial discreteness, but not by a new short-range gravitational interaction. We solve the polymer Schr\"{o}dinger equation in an analytical fashion, and we evaluate numerically the corresponding energy levels. We find that the polymer energy spectrum exhibits a negative shift compared to the obtained for the quantum bouncer. The comparison of our results with those obtained in the GRANIT experiment leads to an upper bound for the fundamental length scale, namely $\lambda \ll 0.6 \buildrel _{\circ} \over {\mathrm{A}}$. We find polymer corrections to the probability of transitions between levels, induced by small vibrations, together with the probability of spontaneous emission in the quadrupole approximation.
\end{abstract}

\pacs{03.65.-w, 04.60.Pp, 03.65.Ge}
\maketitle

\section{Introduction}

One of the main challenges in physics today is the search for a Quantum Theory of Gravity (QTG). On the theoretical front, one of the major problems is that the introduction of gravity into quantum field theories appears to spoil their renormalizability, and from the experimental point of view, the major difficulty is the lack of experimentally accesible phenomena that could shed light on a possible route to QTG. This situation gives rise to the possibility of existence of other fundamental interactions in nature \cite{Yao}, thus providing a strong motivation to the study of gravity at short ranges. The Newton's law has never been tested at distances below of $0.1 \mbox{mm}$, thus opening the possibility of studying extra finite range (non-Newtonian) gravitational interactions. In this scenario different phenomenological approaches have been considered , for example extra Yukawa-type forces \cite{Nesvizhevsky}. Theories with large extra spatial dimensions have also been studied. A proposal consistent with string theory is due to Arkani-Hamed, Dimoupoulos and Dvali (ADD), in which gravity departs from Newton's inverse square law at scales which could be as large as a millimeter \cite{Arkani}. ADD conjectures the existence of two or more additional dimensions in which gravity, but not the strong or electroweak forces, might be acting, diluting itself by spreading its lines of force into these extra dimensions. Essentially, this would explain the apparent weakness of gravity.

Considering the progress that has been achieved recently in neutron physics experiments, it has been proposed that they can reveal deviations from Newton's law at short distances. For example, Nesvizhevsky and co-workers use the best experimental measurements performed with neutron scattering experiments to give constraints for the hypothetical Yukawa-type force \cite{Nesvizhevsky}, and on the other hand, Frank and co-workers propose that slow neutron scattering off atomic nuclei with null spin may provide an experimental test for large extra spatial dimensional gravity \cite{Frank}, but no experiment in this regard has been performed so far. Note that in these works the authors study the quantum effects caused by an extra short-range gravitational potential.

In this paper we adopt a more fundamental point of view. The high energy behaviour of quantum fields is intimately connected with the structure of spacetime at short distances. For example, Loop Quantum Gravity (LQG) asserts that continuous classical spacetime is replaced by quantum spin networks on small scales. Polymer Quantization (PQ) is a background independent quantization scheme that is deployed in LQG. In this programme the notion of discreteness is built-in, and consequently  the momentum operator $p$ is not realized directly as in Schr\"{o}dinger quantum mechanics, but arises indirectly through the translation operator $\widehat{U} _{\lambda} = e ^{- i \frac{p \lambda}{\hbar}}$. PQ has been used to study quantum gravitational corrections to temporal dynamics \cite{MartinDIT,MartinTunn} and to statistical thermodynamics \cite{Chacon} of simple quantum systems. In this paper we use PQ to study the problem of a particle bouncing on a perfectly reflecting surface under the influence of Earth's gravitational field (described by Newton's law), that we have called \textit{The Polymer Bouncer}. In this scenario, deviations from the usual quantum effects are induced by the spatial discreteness, but not by a new short-range gravitational interaction.

Recent high-sensitivity experiments with thermal neutrons free falling in the gravitational field, called GRANIT, demonstrate that the energy spectrum in the gravitational field's direction is quantized \cite{Nesvizhevsky2}. This kind of experiments may offer an opportunity to confront observation with signatures of quantum gravitational effects. To this end in \cref{PolymerBouncer} we solve the problem of the polymer bouncer in an analytical fashion, and we evaluate numerically the corresponding energy levels. The Polymer-Schr\"{o}dinger transition is discussed in \cref{PStransition}. In \cref{GRANIT} we predict an upper bound to the length scale of the polymer theory by confronting our results with the maximal precision of the GRANIT experiment. In \cref{SmallVibrations,TransitionRate} we find polymer corrections for both, the probability of transitions between levels induced by small vibrations and the probability of spontaneous emission in the quadrupole approximation, respectively. Finally, our conclusions are given in \cref{Conclusions}.

\section{The Polymer Bouncer} \label{PolymerBouncer}

We shall consider the problem of a polymer particle of mass $m$ bouncing on a perfectly reflecting surface (at $z = 0$) under the influence of Earth's gravitational field: $\vec{g} = - g \vec{e} _{z}$. For the purposes of this work we will make the following assumptions. Firstly we consider that it is reasonable to keep a constant value for $g$ because the small size of the experiments \cite{Brau}, and secondly we assume that the Weak Equivalence Principle is valid. To tackle this problem, we restrict the dynamics to an equispaced lattice $\gamma \left( \lambda \right) = \left\lbrace \lambda n \; \vert \; n \in \mathbb{Z} ^{+} \right\rbrace$. The spectrum of the position operator, $\left\lbrace z _{n} = \lambda n \right\rbrace$, consists of a countable selection of points from the real line, which is analogous to the graph covering 3-manifolds in LQG. Here $\lambda$ is regarded as a fundamental length scale of the theory.

As in the standard case, the potential for this problem is defined by
\begin{equation}
V _{\mu} = \left\lbrace \begin{array}{c} m g \lambda \mu \;\;\ , \;\;\;\;\;\ \mu > 0, \\ \infty  \;\;\;\;\;\;\;\;\  , \;\;\;\;\;\ \mu < 0, \end{array} \right. . \label{Potential}
\end{equation}
The wave function $\psi _{\mu} \equiv \psi \left( z _{\mu} \right)$ of a polymer bouncer obeys the stationary polymer Schr\"{o}dinger equation for the vertical motion along vertical axis
\begin{equation}
2 \varepsilon \psi _{\mu} =  2 \psi _{\mu} - \psi _{\mu + 1} - \psi _{\mu - 1} + \upsilon ^{-1} \mu \psi _{\mu} . \label{PolSchrEq}
\end{equation}
where we have defined the following dimensionless quantities
\begin{equation}
\varepsilon = \frac{m \lambda ^{2} E}{\hbar ^{2}} \;\;\;\;\;\ , \;\;\;\;\;\ \upsilon = \left( l _{0} / \lambda \right) ^{3} , \label{Dimensionless} 
\end{equation}
with $l _{0} = \left( \frac{\hbar ^{2}}{2 m ^{2} g} \right) ^{1/3}$ the characteristic gravitational length \cite{Gibbs}. The appropriate solution for this problem is realized through the following boundary conditions: $\psi _{\mu}$ must vanish asymptotically as $\mu \rightarrow \infty$, and $\psi _{0} = 0$ because of the mirror at $\mu = 0$.

In most cases, second-order linear difference equations with variable coefficients cannot be solved in closed form. Fortunately, for the special case of the difference equation (\ref{PolSchrEq}) it is possible to obtain an analytic solution in a simple fashion \cite{Ehrhardta}. The most convenient approach for finding the solution of difference equations with polynomial coefficients is the method of the generating function. Let
\begin{equation}
f \left( \zeta \right) = \sum _{\mu = - \infty} ^{+ \infty} \psi _{\mu} \zeta ^{\mu} , \label{GeneratingF}
\end{equation}
be the generating function for the solution of \cref{PolSchrEq}. Note that (\ref{GeneratingF}) is not a generating function in the strict sense but a Laurent series, which is uniformly convergent, and thus we can differentiate each term. Multiplying \cref{PolSchrEq} by $\zeta ^{\mu -1}$ and summing for $\mu \in \mathbb{Z}$, the following differential equation for the generating function is obtained
\begin{equation}
\left[ \frac{d}{d \zeta} - \upsilon \frac{\zeta ^{2} - 2 \left( 1 - \varepsilon \right) \zeta + 1}{\zeta ^{2}} \right] f \left( \zeta \right) = 0 , \label{DiffEqF}
\end{equation}
for which the solution is
\begin{equation}
f \left( \zeta \right) = \zeta ^{-2 \upsilon \left( 1 - \varepsilon \right)} e ^{\upsilon \left( \zeta - \frac{1}{\zeta} \right)} = \zeta ^{-2 \upsilon \left( 1 - \varepsilon \right)} \sum _{\mu = - \infty} ^{+ \infty} J _{\mu} \left( 2 \upsilon \right) \zeta ^{\mu} , \label{SolGeneratingF}
\end{equation}
where we have identified the generating function for the Bessel functions of the first kind $J _{n}$. After relabeling the index in \cref{SolGeneratingF}, a simple comparison with \cref{GeneratingF} shows that the decaying solution of \cref{PolSchrEq} is
\begin{equation}
\psi _{\mu} = N_{\varepsilon} J _{\mu + 2 \upsilon \left( 1 - \varepsilon \right)} \left( 2 \upsilon \right) , \label{WaveFunc}
\end{equation}
where $N_{\varepsilon}$ is the appropriate normalization factor. By using the recurrence relation for Bessel functions, $J _{\nu + 1 } \left( z \right) + J _{\nu - 1 } \left( z \right) = 2 \frac{\nu}{z} J _{\nu} \left( z \right)$, one can further check that \cref{WaveFunc} correctly solves the time-independent polymer Schr\"{o}dinger equation (\ref{PolSchrEq}). This generating function approach is not suitable for determining the growing solution of \cref{PolSchrEq}, but it is well known that the Neumann function $Y _{\mu}$ also satisfies the recursion equation for Bessel functions. It does not however, represent a physical solution for this problem.

The second boundary condition, $\psi _{0} = 0$, implies that $2 \upsilon$ must be chosen as a root of the Bessel function $J _{2 \upsilon \left( 1 - \varepsilon \right)} \left( 2 \upsilon \right) = 0$, where $2 \upsilon \left( 1 - \varepsilon \right) \in \mathbb{R}$ (not necessarily an integer). Let $j _{[n,r]}$ be the $r$-th zero of the Bessel function $J _{n}$. Then the condition
\begin{equation}
2 \upsilon = j _{[2 \upsilon \left( 1 - \varepsilon \right),n]} \label{PolEnergySpec}
\end{equation}
defines the quantized energy levels of the polymer bouncer as a function of $\upsilon$. Unfortunately \cref{PolEnergySpec} cannot be solved analitically, but it can be studied numerically for arbitrary values of $\upsilon$. Using the quantization condition (\ref{PolEnergySpec}), we can compute the normalization factor in a simple fashion. One obtains
\begin{equation}
\sum _{\mu = 0 } ^{\infty} \vert \psi _{\mu} \vert ^{2} = N_{\varepsilon} ^{2} \upsilon J _{1 + 2 \upsilon \left( 1 - \varepsilon \right)} \left( 2 \upsilon \right) J ^{(1,0)} _{ \upsilon \left( 1 - \varepsilon \right)} \left( 2 \upsilon \right) = 1 , \label{Normalization}
\end{equation}
where $J ^{(1,0)} _{\alpha} \left( z \right) = \frac{\partial J _{\alpha ^{\prime}} \left( z \right)}{\partial \alpha ^{\prime}} \Big| _{\alpha ^{\prime} = \alpha} $ is the derivative of the Bessel function with respect to its order. The substitution of $N_{\varepsilon}$ into the polymer wave function \cref{WaveFunc} establishes the normalized polymer wave function. Now we proceed with the analysis of the polymer energy spectrum.

For the sake of simplicity, let us assume that the gravitational length $l _{0}$ is a multiple of the fundamental length $\lambda$, i.e. $l _{0} = s \lambda $ with $s \in \mathbb{Z} ^{+}$. In \cref{table} we present the first ten energy levels of the polymer bouncer as a function of $s$. We point out that these numerical estimates are good approximations for $ s > 1 $. The case $s = 1$ exhibits difficulties arising from the polymer behaviour, which we discuss later. Our results show that the polymer energy levels display a negative shift with respect to the energy levels of the quantum bouncer. Such negative shift in energy can be understood by studying the dynamics of the polymer bouncer. The Heisenberg equation of motion for the operator $\hat{p} _{\lambda}$ yields
\begin{equation}
\frac{d \hat{p} _{\lambda}}{dt} = \frac{i}{\hbar} m g \left[ \hat{z} , \hat{p} _{\lambda} \right] = - m g \cos \frac{p \lambda}{\hbar} \geq - m g \label{HeisEqMot}
\end{equation}
where the inequality is established using the domain of the momentum in the polymer case, i.e. $p \in \left[ - \pi \hbar / \lambda , \pi \lambda / \hbar \right] $. In the first order of approximation, \cref{HeisEqMot} becomes $\dot{\hat{p}} _{\lambda} \approx - m g \left( 1 - \frac{p ^{2} \lambda ^{2}}{2 \hbar} \right)$. So, clasically, an additional $p ^{2}$ dependent force acts on the particle which decreases the energy of the system.

Quantum mechanically, the negative shift is in agreement with the correction term of the regularized $\hat{p} ^{2} _{\lambda}$ operator in the polymer theory for $p \lambda \ll \hbar $, i.e.
\begin{equation}
\hat{p} ^{2} _{\lambda} \approx p ^{2} - \frac{\lambda ^{2}}{12 \hbar ^{2}} p ^{4} . \label{RegP2Op}
\end{equation}
With the assumption that the fundamental length is very small compared with the characteristic gravitational length, $\lambda \ll l _{0}$, the energy shift can be roughly estimated. In first order perturbation theory, the enery shift becomes
\begin{equation}
\Delta \varepsilon _{n} = - \frac{\lambda ^{4}}{24 \hbar ^{4}} \left< \psi _{n} \right| p ^{4} \left| \psi _{n} \right> = - \frac{m ^{2} \lambda ^{4}}{6 \hbar ^{4}} \left< \psi _{n} \right| \left( E _{n} - m g z \right) ^{2} \left| \psi _{n} \right> . \label{EnergyShift}
\end{equation}
The averages in this expression can be easily computed by using the quantum-mechanical wave function \cref{WaveEquation1}. The final result is 
\begin{equation}
\Delta \varepsilon _{n} = - \frac{a _{n} ^{2}}{120 s ^{4}} , \label{EnergyShift2}
\end{equation}
which is in a close agreement with the results reported in \cref{table}. Due to the lack of precision in numerical calculations for the case $ s = 1 $ in \cref{PolEnergySpec}, we have employed the energy shift of \cref{EnergyShift2} in \cref{table} only for this case. The energy spectrum of a quantum bouncer also has been studied in the Generalized Uncertainty Principle (GUP) framework \cite{Brau}. In this case the modification induced by a minimal length $l _{\min}$ on the energy spectrum is
\begin{equation}
E ^{(GUP)} _{n} = - m g l _{0} a _{n} + \alpha ^{2} l _{\min} ^{2} a _{n} ^{2} , \label{GUPSpectrum}
\end{equation}
where $\alpha ^{2}$ is a GUP parameter. Note that both the polymer- and the GUP-corrections, depend on the minimum length and the zeros of the Airy's function quadratically, but they feature a sign difference. Such opposite tendency is also present for the energy spectrum for a particle in a box, as pointed out in Ref.\cite{Chacon}. 

Another important consequence of the commutator in \cref{HeisEqMot} is the modification of the position-momentum uncertainty relation. We know that for two operators $\hat{A}$ and $\hat{B}$, the identity $\left( \Delta \hat{A} \right) ^{2} \left( \Delta \hat{B} \right) ^{2} = \vert \left\langle \left[ \hat{A} , \hat{B} \right] \right\rangle \vert ^{2} / 4$ holds irrespective of the quantization scheme \cite{Hossain}. In the polymer framework the position-momentum uncertainty realtion reads
\begin{equation}
\left( \Delta \hat{z} \right) ^{2} \left( \Delta \hat{p} _{\lambda} \right) ^{2} = \frac{\hbar ^{2}}{4} \Big| \left\langle  \cos \left( \frac{p \lambda}{\hbar} \right) \right\rangle  \Big| ^{2} \approx \frac{\hbar ^{2}}{4} \left[ 1 - \frac{\left\langle p ^{2} \right\rangle  \lambda ^{2}}{2 \hbar ^{2}} + \mathcal{O} \left( \lambda ^{4} \right) \right] . \label{UncertaintyRel}
\end{equation}
Note that the correction term in the right hand side is negative definite, thus implying that the uncertainty decreases due to the presence of $\lambda$. \Cref{UncertaintyRel} suggests that polymer quantum mechanics at short enough wave lengths (of the order of $\sim 2 \lambda $) can exhibit classical behaviour. This result has a strong resemblance to the one found, on a different setup, by Magueijo and Smolin in deformed special relativity \cite{Magueijo}. In this framework, 't Hooft has discussed the possibility of a deterministic quantum mechanics at Planck scale, supplemented with a dissipation mechanism, giving rise to the standard quantum mechanical behaviour at larger scales \cite{Hooft}. \Cref{UncertaintyRel} has been used to derive a new mass-temperature relation for Schwarzschild (micro) black holes \cite{Jizba}.

\begin{figure}
\subfloat[Ground state, $s=10$,\label{s10n1}]{\includegraphics[width = 2.5in]{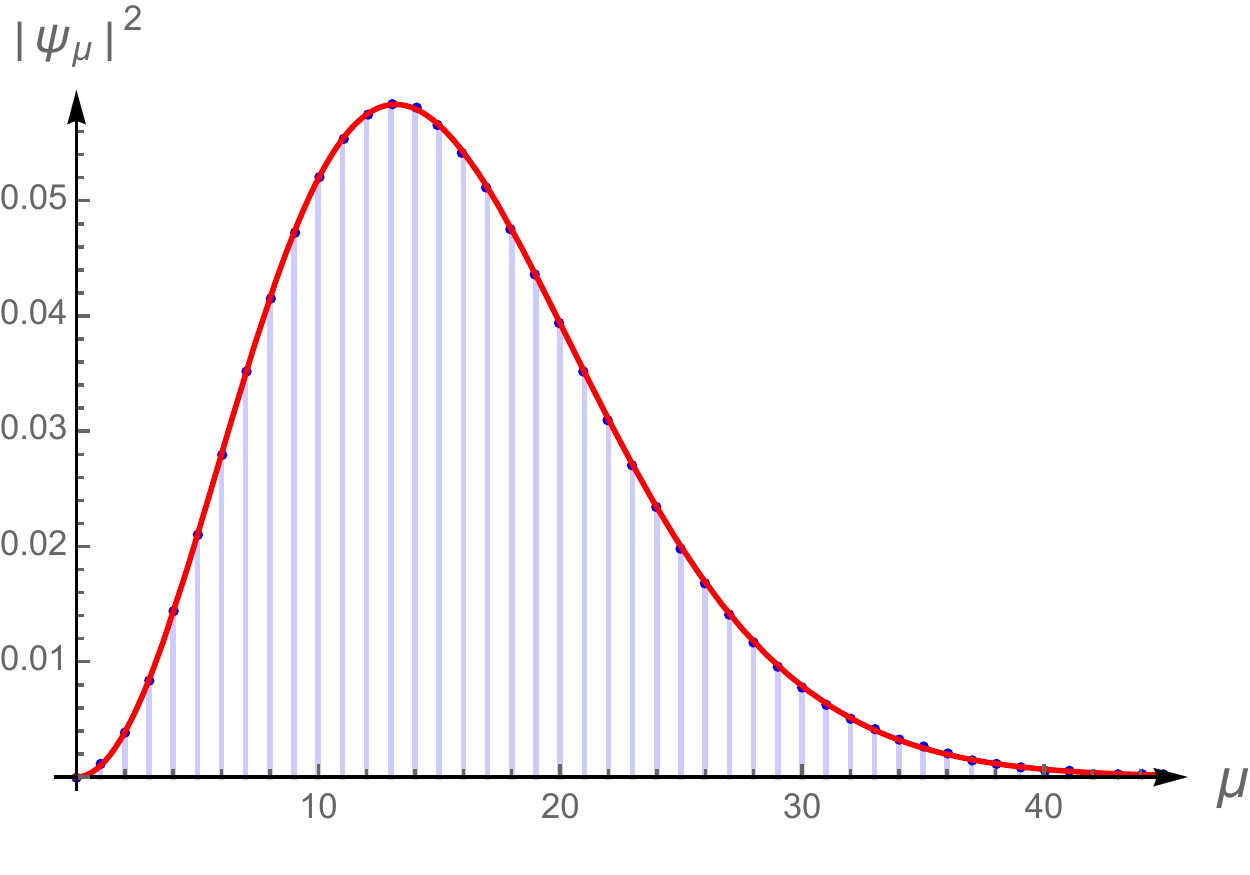}}
\subfloat[Ground state, $s=5$,\label{s5n1}]{\includegraphics[width = 2.5in]{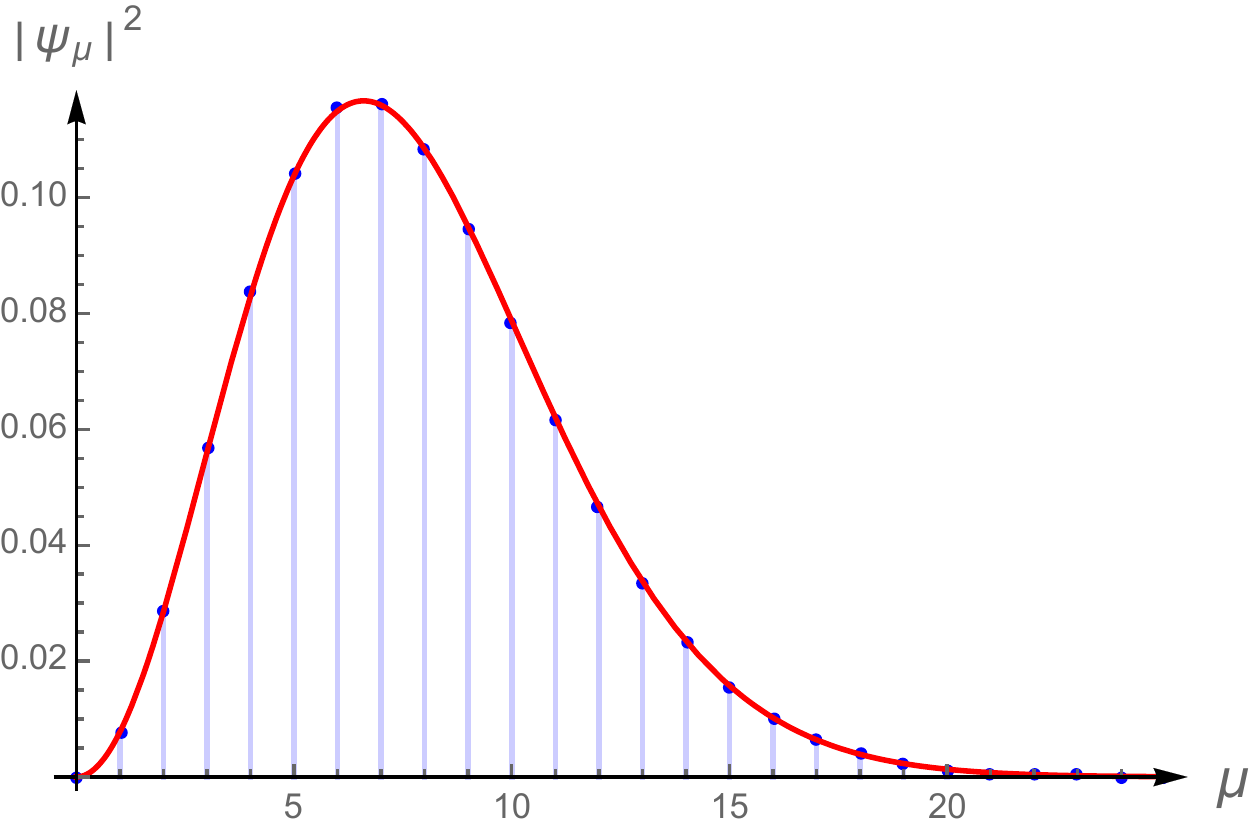}}
\subfloat[Ground state, $s=1$,\label{s1n1}]{\includegraphics[width = 2.5in]{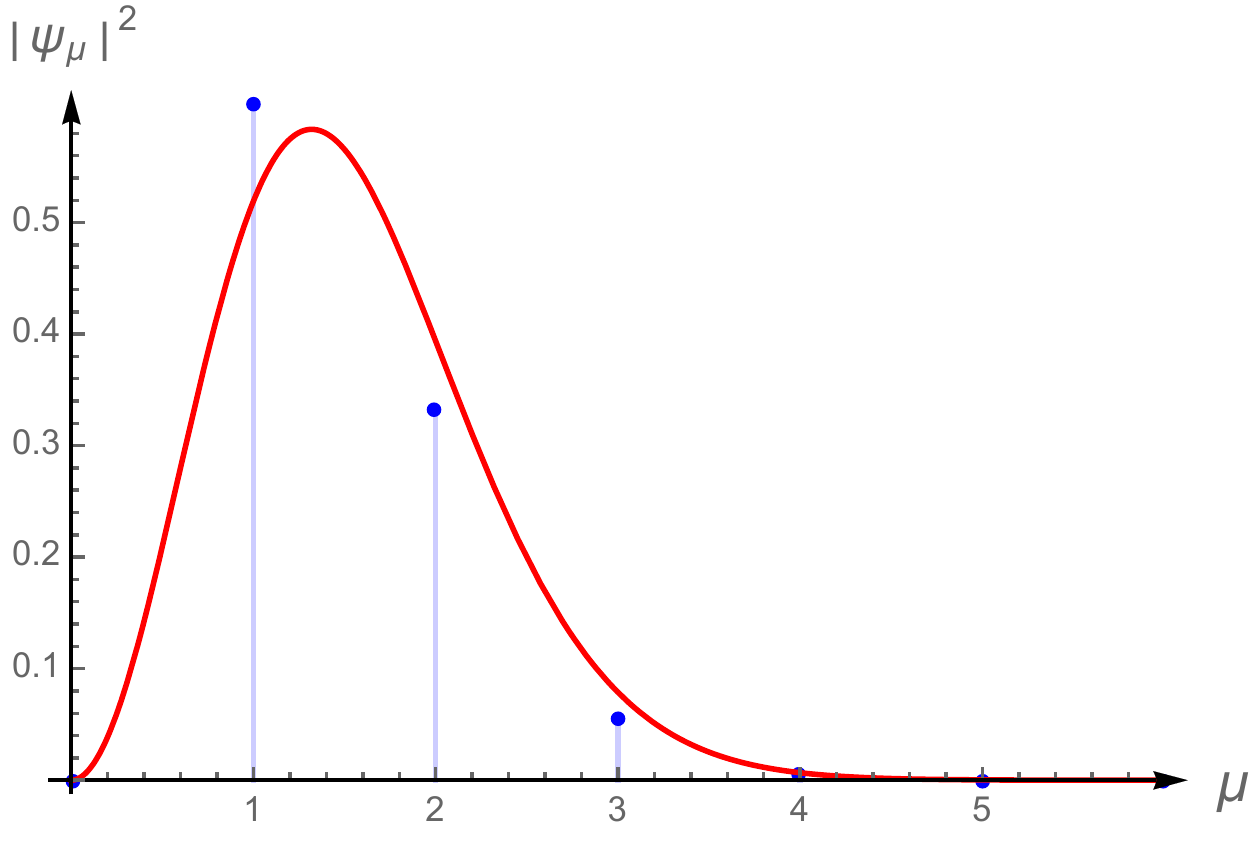}} \\
\subfloat[First excited state, $s=10$,\label{s10n2}]{\includegraphics[width = 2.5in]{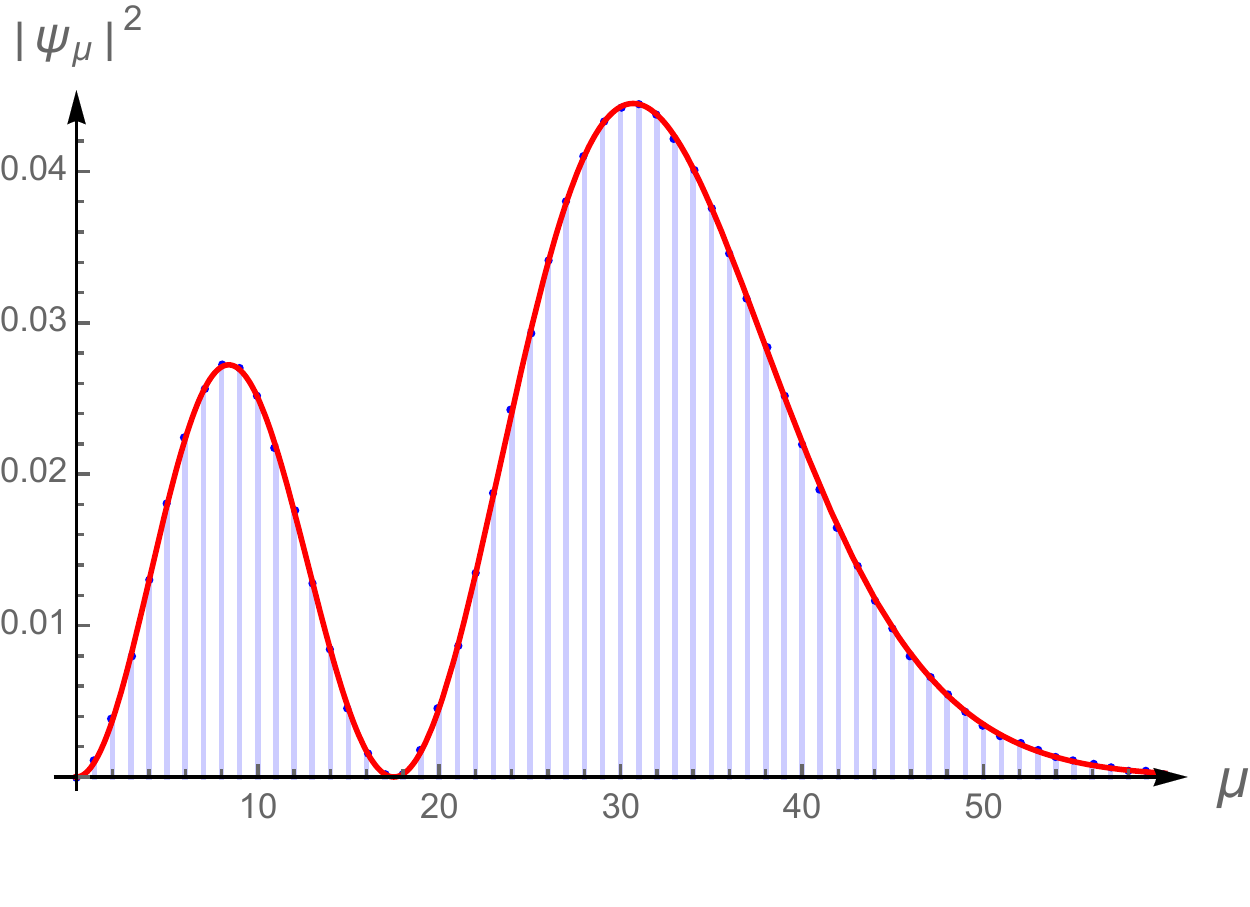}}
\subfloat[First excited state, $s=5$,\label{s5n2}]{\includegraphics[width = 2.5in]{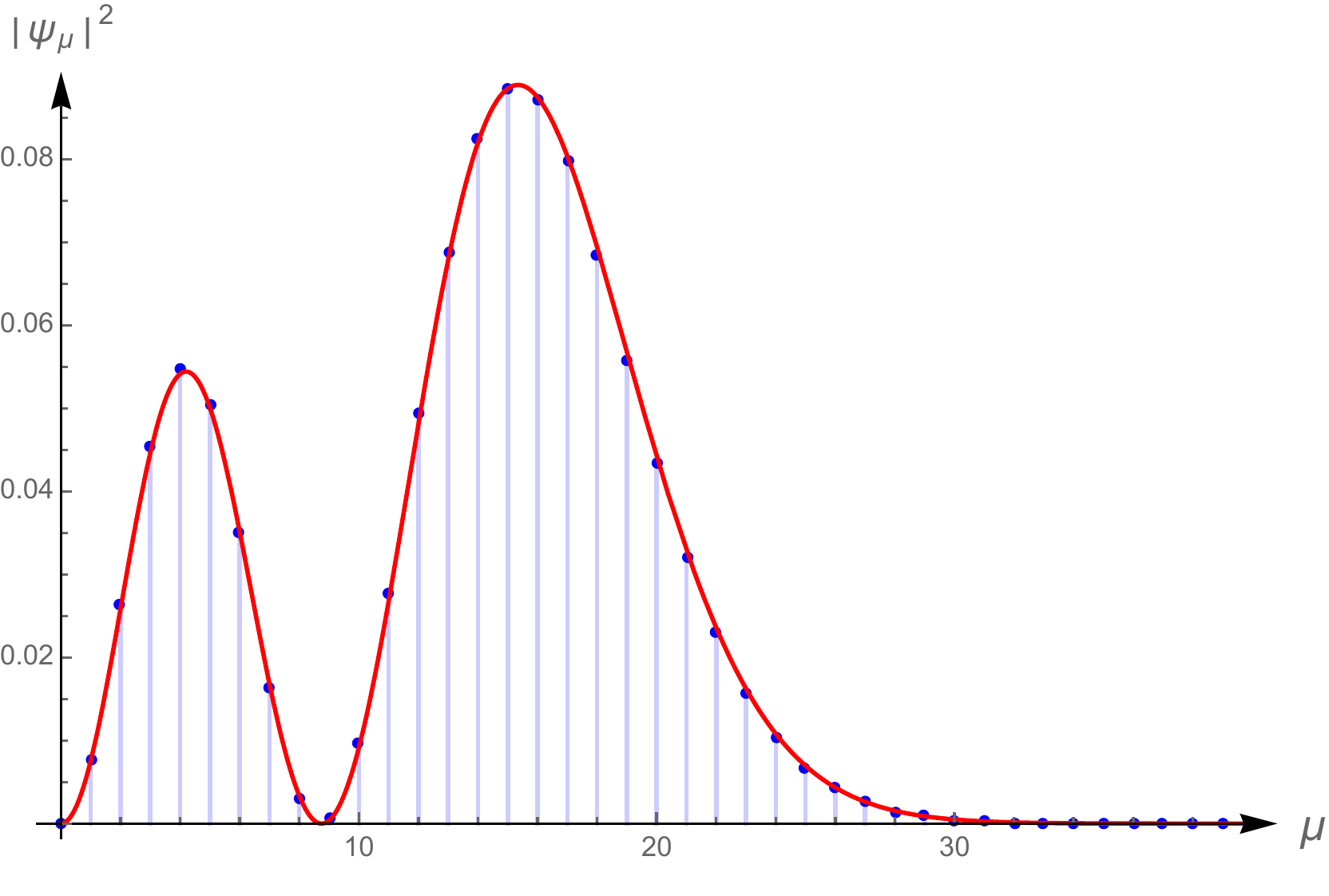}} 
\subfloat[First excited state, $s=1$,\label{s1n2}]{\includegraphics[width = 2.5in]{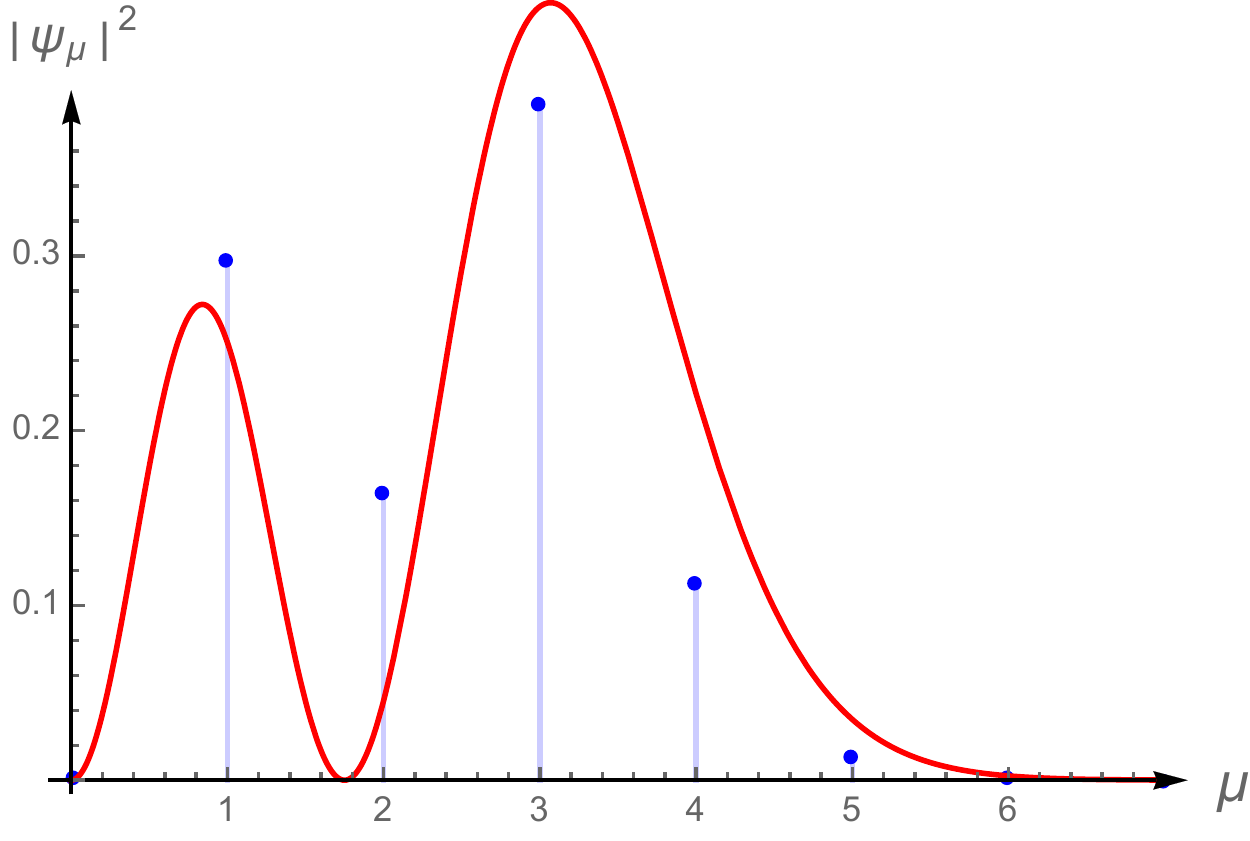}} \\
\subfloat[Tenth excited state, $s=10$,\label{s10n10}]{\includegraphics[width = 2.5in]{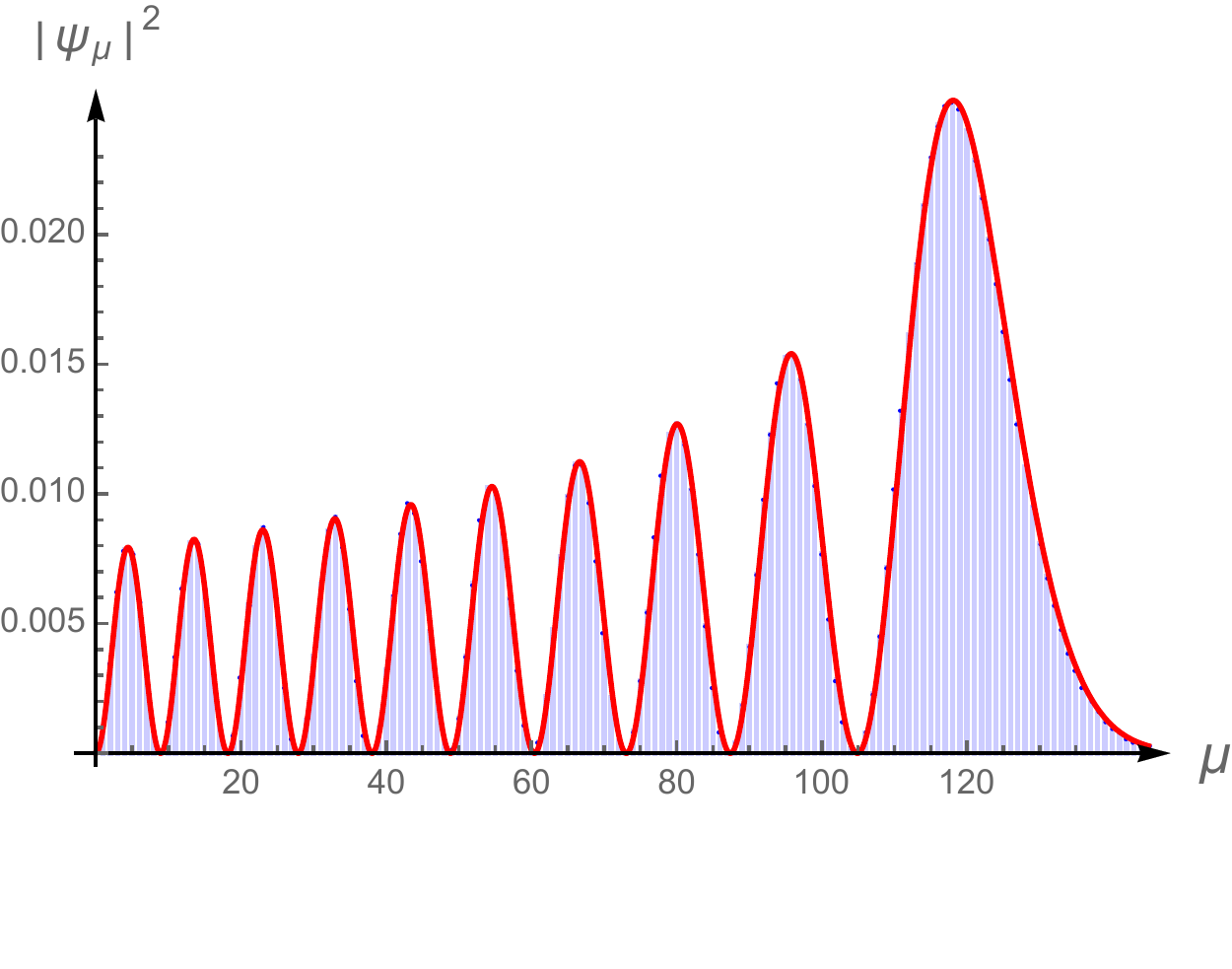}}
\subfloat[Tenth excited state, $s=5$,\label{s5n10}]{\includegraphics[width = 2.5in]{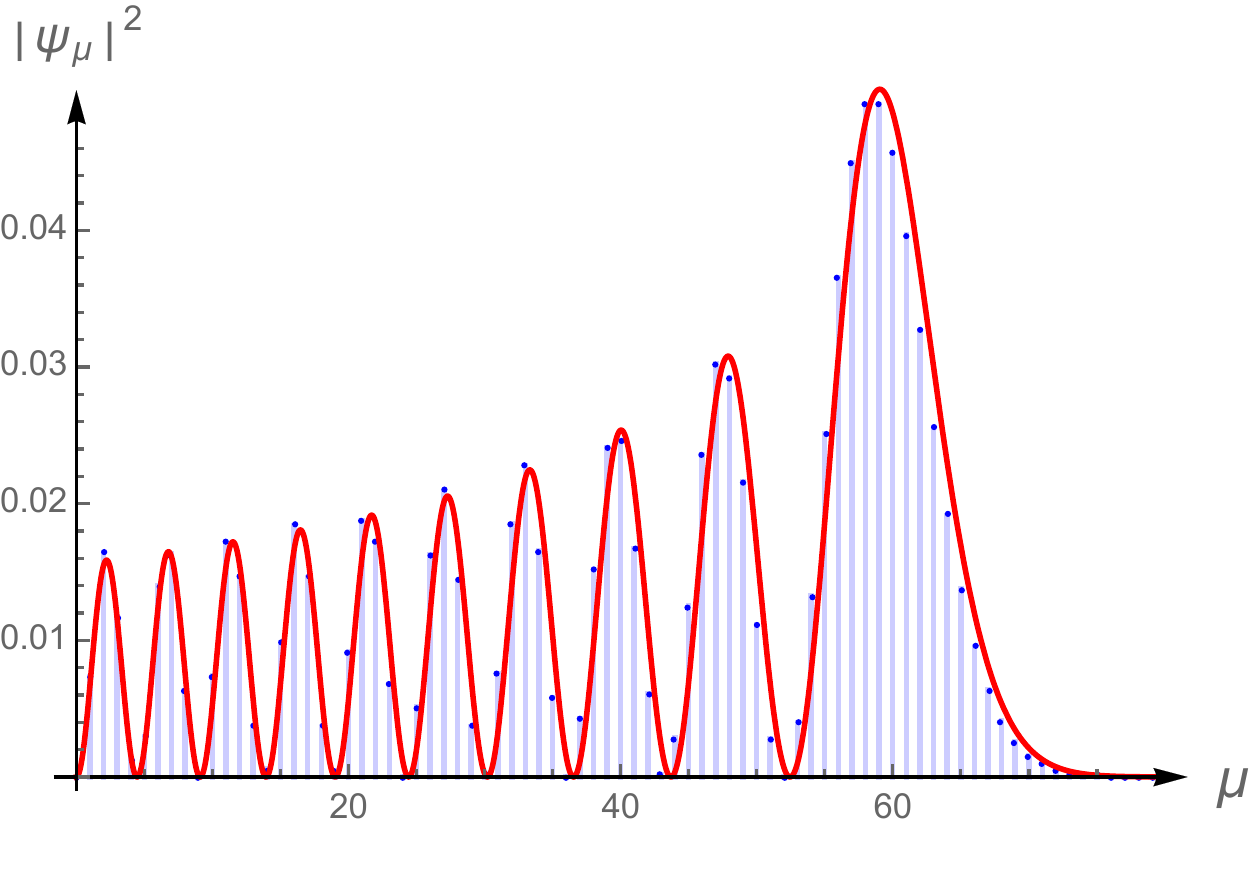}} 
\subfloat[Tenth excited state, $s=2$,\label{s2n10}]{\includegraphics[width = 2.5in]{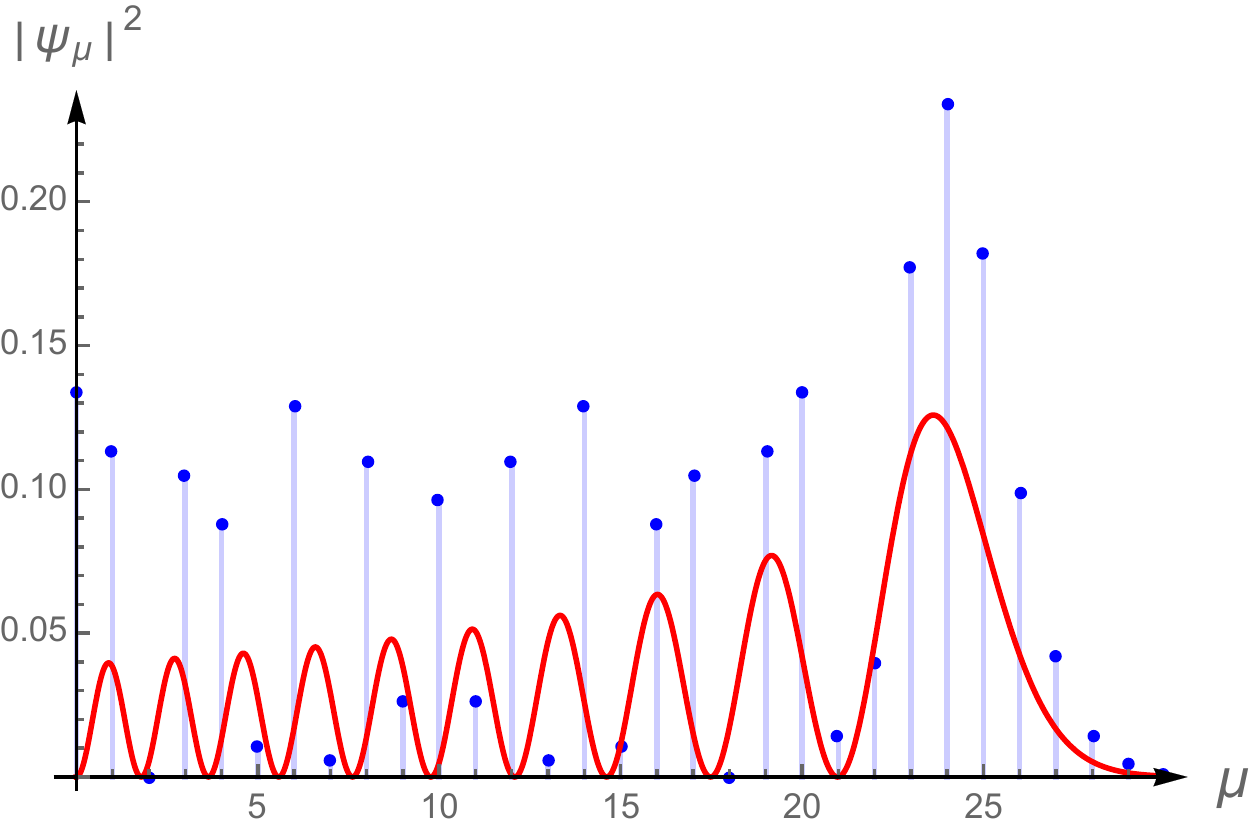}} 
\caption{\small Plots of the polymer (blue dots) and quantum-mechanical (red lines) density profiles for different energy levels and different values of $s$.}
\label{DensityProfiles}
\end{figure}

We conclude this section by comparing the polymer and quantum density profiles for different energy levels and different values of $s$. Hereafter we denote by $\psi _{\mu} ^{(n)}$ the polymer wave function of the $n$-th state. The discrete plot corresponds to the polymer result, and the solid line corresponds to its quantum-mechanical counterpart. \Cref{s10n1,s5n1,s1n1} show the ground state for $s=10,5$ and $1$, respectively. \Cref{s10n2,s5n2,s1n2} show the first excited state for $s=10,5$ and $1$, respectively. \Cref{s10n10,s5n10,s2n10} show the tenth excited state for $s=10,5$ and $2$, respectively. We expect a close agreement between the polymer and quantum-mechanical density profiles for the case $s \gg 1$, however in \cref{DensityProfiles} we observe that this occurs already for $s=5$. When the gravitational length is of the order of the fundamental length ($s = 1$) the polymer distribution departs from its quantum-mechanical counterpart, and the polymer effects become important. From \cref{s2n10} we can infer that the most significant polymer effects are for high energies and $s = 1$, as expected.

\begin{center}
\begin{table}
\scalebox{0.7}{
\begin{tabular}{ | c | c | c | c | c | c |  c | c | c | c | c | c |}
  \hline       
   \diagbox{s}{n} & 1 & 2 & 3 & 4 & 5 & 6 & 7 & 8 & 9 & 10 \\ \hline              
  10 & 0.011686 & 0.0204258 & 0.0275773 & 0.033895 & 0.0396679 & 0.0450452 & 0.0501165  & 0.0549412 & 0.0595607 & 0.064006 \\          
   9 & 0.0144258 & 0.025213 & 0.0340387 & 0.0418346 & 0.0489574 & 0.0555915 & 0.0618477 & 0.067799 & 0.073497 & 0.0789795 \\ 
   8 & 0.0182553 & 0.031903 & 0.0430672 & 0.052927 & 0.0619345 & 0.0703228 & 0.0782324 & 0.0857557 & 0.0929579 & 0.0998871 \\
   7 & 0.0238393 & 0.0416556  & 0.056226 & 0.0690913 & 0.080842 & 0.0917831 & 0.102098 & 0.111907 & 0.121296 & 0.130328 \\   
   6 & 0.0324385 & 0.0566692 & 0.0764773 & 0.0939613 & 0.109926 & 0.124785 & 0.138791 & 0.152107 & 0.164849 & 0.177103 \\ 
   5 & 0.0466892 & 0.0815348 & 0.110001 & 0.135113 & 0.15803 & 0.17935 & 0.199436 & 0.218523 & 0.23678 & 0.254331 \\ 
   4 & 0.0728877 & 0.127199 & 0.171511 & 0.210558 & 0.246155 & 0.279241 & 0.310382 & 0.339949 & 0.368205 & 0.395345 \\ 
   3 & 0.129331 & 0.22536 & 0.303481 & 0.372143 & 0.434588 & 0.492495 & 0.546873 & 0.59839 & 0.647516 & 0.694599 \\
   2 & 0.289409 & 0.501951 & 0.673219 & 0.822395 & 0.956849 & 1.1875 & 1.3125 & 1.4375 & 1.5625 & 1.625 \\   
   1 & 1.1235 & 1.90471& 2.50631& 3.00953 & 3.44616 & 3.83292 & 4.18004 & 4.49437 & 4.78077 & 5.04291  \\ \hline 
\end{tabular}
}
\caption{\small Estimates of the first ten (rescaled) energy levels of the polymer bouncer as a function of $s$.} \label{table}
\end{table}
\end{center}

\section{The Polymer-Schr\"{o}dinger transition} \label{PStransition}

In physics, if a new theory is considered more general that the former, it must yield accurate results not only in the new scale but at the former one as well. In this sense, we know that Newtonian mechanics can be recovered from relativistic mechanics in the domain of low velocities compared with the speed of light in vacuum. Regarding the quantum-classical transition the problem is more subtle, given that the conceptual framework of these theories are fundamentally different \cite{Martin,Bernal}. In the problem at hand one expects that if the lattice spacing $\lambda$ is taken to be sufficiently small, the polymer formulation should reduce to the Schr\"{o}dinger representation. However, this is a delicate issue because $\lambda$ is regarded as a nonzero fundamental length scale of the polymer theory, and it cannot be removed when working in the polymer Hilbert space $H _{poly}$ \footnote{The kinemathical Hilbert space can be written as $H _{poly} = L ^{2} \left( \mathbb{R} _{d} , d \mu _{d} \right)$, with $d \mu _{d}$ corresponding Haar measure, and $\mathbb{R} _{d}$ the real line endowed with a discrete topology}, no matter how small $\lambda$ is \cite{MartinDIT}.

The Polymer-Schr\"{o}dinger transition for this problem is quite simple. Taking $\lambda$ to be of the order of the Planck length, and the value of the characteristic length for the falling slow neutrons case $l _{0} = 5.87 \mu$m \cite{Codau}, one obtains $\upsilon \sim 10 ^{88} $. Note that this approximation is valid even if we take $\lambda$ several orders of magnitude larger than the Planck lenght. Therefore, the asymptotic behaviour of the Bessel functions for large arguments is required in \cref{WaveFunc}. On the other hand, the $l _{0} \gg \lambda$ limit also implies that we should take a very large number of points between two arbitrary points, i.e. $\mu \gg 1$. Therefore $\mu + 2 \upsilon \left( 1 - \varepsilon _{n} \right) > 2 \upsilon \gg 1$, and the asymptotic behaviour of the Bessel function for large orders is also required. The asymptotic expansion for the Bessel functions for these conditions is well known \cite{Gradshteyn}. For $n \gg 1$ and $n > x$ the following approximation is valid
\begin{equation}
J _{n} \left( x \right) \sim \frac{1}{3} \sqrt{\frac{2 \left( x - n \right)}{x}} \left\lbrace J _{1/3} \left[\frac{\left\lbrace 2 \left( x - n \right) \right\rbrace ^{2/3}}{3 \sqrt{x}} \right] + J _{-1/3} \left[\frac{\left\lbrace 2 \left( x - n \right) \right\rbrace ^{2/3}}{3 \sqrt{x}} \right] \right\rbrace . \label{Asymptotics1}
\end{equation}
Now by using the identity of \cref{Airy-Bessel} we find
\begin{equation}
J _{n} \left( x \right) \sim \left( \frac{2}{x} \right) ^{1/3} Ai \left[ \left( \frac{2}{x} \right) ^{1/3} \left( n - x \right) \right] , \label{Asymptotics2}
\end{equation}
and by setting $n = \mu + 2 \upsilon \left( 1 - \varepsilon _{n} \right) \gg 1 $ and $x = 2 \upsilon \gg 1 $ we obtain
\begin{equation}
\psi _{\mu} ^{(n)} \sim \frac{N _{n}}{\upsilon ^{1/3}} Ai \left[\frac{z}{l _{0}} - \frac{E _{n}}{mgl _{0}} \right] , \label{Asymptotics3}
\end{equation}
where $z = \lambda \mu$ and $E _{n}$ is defined through \cref{Dimensionless}. From this expression we also recover the quantum-mechanical energy spectrum as proportional to the zeros of Airy functions.

On the other hand, the normalization factor can be approximated by using the Euler-Maclaurin formula,
\begin{equation}
\sum _{\mu = 0} ^{\infty} \vert \psi _{\mu} ^{(n)} \vert \sim \frac{N _{n} ^{2}}{\upsilon ^{2/3}} \int _{0} ^{\infty} d \mu Ai ^{2} \left[ \frac{\mu \lambda}{l _{0}} - \frac{E _{n}}{mg l _{0}} \right] =  \frac{N _{n} ^{2}}{l _{0}} Ai ^{\prime 2} \left[ - \frac{E _{n}}{mg l _{0}} \right] = 1 , \label{Asymptotics4}
\end{equation}
where we have used that $Ai \left( x \right) \rightarrow 0 $ as $x \rightarrow \infty $ and $Ai \left[ - \frac{E _{n}}{mg l _{0}} \right] = 0$. Substituting this result into \cref{Asymptotics3} we finally obtain
\begin{equation}
\frac{\psi _{\mu} ^{(n)}}{\sqrt{\lambda}} \sim \psi _{n} \left( z \right) , \label{Asymptotics5}
\end{equation}
where $\psi _{n} \left( z \right)$ is the wave function for the quantum bouncer \cref{WaveEquation1}. Note that the presence of $\lambda ^{-1/2}$ is consistent with the fact that the Lebesgue measure has dimensions of length while the discrete measure is dimensionless \cite{Flores}.

\section{Comparison with the GRANIT experiment} \label{GRANIT}

The experimental physics of slow neutrons has undergone significant evolution in the last decades. Recent high-sensitivity experiments, called GRANIT, performed by V. V. Nesvizhevsky \textit{et al.} at the \textit{Institute Laue-Langevin}, show that cold neutrons moving in a gravitational field do not move smoothly but jump from one height to another, as predicted by quantum theory \cite{Nesvizhevsky2}. They used an intense horizontal beam of cold neutrons directed slightly upwards and allowing the neutrons to fall onto a horizontal mirror. By placing a neutron absorber above the mirror and counting the particles as they moved the absorber up and down, they found that neutrons are measured only at certain well-defined heights. The experimental average values of the two lowest critical heights (taken from \cite{Nesvizhevsky3}) are
\begin{eqnarray}
h _{1} ^{\mbox{exp}} &=& \left( 12.2 \pm 1.8 _{sys} \pm 0.7 _{stat} \right) \mu \mbox{m} , \label{HeightsExp} \\ h _{2} ^{\mbox{exp}} &=& \left( 21.6 \pm 2.2 _{sys} \pm 0.7 _{stat} \right) \mu \mbox{m} , \nonumber
\end{eqnarray}
while the theoretical heights ($h _{n} = - a _{n} l _{0}$) with $m = 1.67 \times 10 ^{27}\mbox{Kg}$ and $g = 9.806 \mbox{m}/\mbox{s} ^{2}$ gives
\begin{equation}
h _{1} = 13.7 \mu \mbox{m} \;\;\;\;\;\ , \;\;\;\;\;\ h _{2} = 24.0 \mu \mbox{m} . \label{HeightsTheo}
\end{equation}
The theoretical values are therefore located within the error bars. As a consequence of the good agreement between theory and experiment, this finding could be used for bounding deviations from the standard theory due to an eventual new physical mechanism. In the problem at hand, the (negative) energy shift must satisfy the constraint
\begin{equation}
\vert \Delta E _{n} \vert < \Delta E _{n} ^{\mbox{exp}} , \label{BoundEnergy}
\end{equation}
with $\Delta E _{n}$ the energy shift given by \cref{EnergyShift2} and $\Delta E _{n} ^{\mbox{exp}}$ the maximal experimental error. The substitution of \cref{EnergyShift2} into \cref{BoundEnergy} produces an upper bound for $\lambda$ given by
\begin{equation}
\lambda ^{2} < \frac{60 l _{0}}{m g a _{n} ^{2}} \Delta E ^{\mbox{exp}} _{n} . \label{BoundLambda}
\end{equation}
Let us discuss the possible bounds for $\lambda$. For the first two states, we know that $\Delta E _{1} ^{\mbox{exp}} = 0.102 \mbox{peV}$ and $\Delta E _{2} ^{\mbox{exp}} = 0.051\mbox{peV}$ \cite{Nesvizhevsky3}. With this values \cref{BoundLambda} yields $\lambda < 10 ^{-6} \mbox{m}$. This result tells us that the effects of the spatial discretization would be largely unobservable in the GRANIT experiment, even with an improvement of the experimental precision. On the other hand \cref{BoundLambda} suggests that a better upper bound can be established if the gravitational field is intensified by some mechanism. Following this idea, Nesvizhevsky and co-workers have considered the centrifugal states of neutrons, which is the quantum analog of the so-called whispering gallery wave \cite{Nesvizhevsky4}. They consider the scattering of cold neutrons by a perfect cylindrical mirror with a radius of a few centimeters. They found that neutrons are affected by a huge centrifugal accelerations of the order $10^{5} - 10^{7} g$. Most neutrons entering at a tangential trajectory are deviated to small angles. However, some neutrons could be captured into long-living centrifugal states  \cite{Nesvizhevsky5}. In the limiting case we are considering here ($\lambda \ll l _{0}$) our results can be applied to the centrifugal quasistationary states, but not in the polymer regime ($\lambda \sim l _{0}$), because of the (spontaneous or) explicit breaking of continuous symmetries \cite{Ghanashyam}. The implementation of this centrifugal acceleration into \cref{BoundLambda} yields $\lambda < 0.6 \buildrel _{\circ} \over {\mathrm{A}}$, which in turn is greater than the mean square neutron radius $r_{n} \sim 0.8 \; \mbox{fm}$, thus implying that actually this limit should be read as $\lambda \ll 0.6 \buildrel _{\circ} \over {\mathrm{A}}$. This is the best bound we can establish on the fundamental length of the polymer theory, with respect to the best measurements performed in the GRANIT experiment.

\section{Transitions Induced By Small Vibrations} \label{SmallVibrations}

It is commonly believed that $\lambda$ is in the order of the Planck length $ \sim 10 ^{-35} \mbox{m}$ (no known data substantiates this conjecture), which is $10 ^{24}$ times smaller than the upper bound we established in the previous section. Therefore our results are not yet feasible for the energy scales of present experiments, but they motivates the search of other mechanisms that could provide a better bounds for the fundamental length.

Although the quantum states of the bouncing neutron are fundamentally stable, various physical effects can induce transitions. In this framework, vibrations and waviness of the mirror surface have been analyzed in Ref.\cite{Codau}. The purpose of this section is to study whether the polymer quantization scheme induces corrections to the probability of transition between states that could be significant for detecting quantum gravitational effects.

Let us assume that the mirror vibrates with a time-dependent height described by a time-dependent function $h \left( t \right) $ with support on the equispaced lattice $\gamma \left( \lambda \right) = \left\lbrace \lambda n \vert n \in \mathbb{Z} ^{+} \right\rbrace $. By performing the transformation
\begin{equation}
\mu ^{\prime} = \mu - h \left( t \right) \;\;\;\;\ , \;\;\;\;\ \phi _{\mu ^{\prime}} \left( t \right) = \psi _{\mu} \left( t \right) , \label{Transformation}
\end{equation}
the modified polymer Schr\"{o}dinger equation becomes
\begin{equation}
i \hbar \frac{\partial}{\partial t} \phi _{\mu ^{\prime}} = - \frac{\hbar ^{2}}{2 m \lambda ^{2}} \left( \phi _{\mu ^{\prime} + 1} - 2 \phi _{\mu ^{\prime}} + \phi _{\mu ^{\prime} +1} \right) + m g \lambda \mu ^{\prime} \phi _{\mu ^{\prime}} + \left[ m g \lambda h + i \dot{h} \frac{\hbar}{2 \lambda} \left( \widehat{U} _{\lambda} - \widehat{U} _{- \lambda} \right) \right] \phi _{\mu ^{\prime}} , \label{ModPolSchrEq}
\end{equation}
where $\dot{h}$ is the time-derivative and $\widehat{U} _{\lambda}$ the translation operator. We observe that this equation consists of the standard polymer Schr\"{o}dinger equation for the polymer bouncer (\ref{PolSchrEq}) plus an effective perturbation potential
\begin{equation}
\widehat{V} = m g \lambda h + i \dot{h} \frac{\hbar}{2 \lambda} \left( \widehat{U} _{\lambda} - \widehat{U} _{- \lambda} \right) . \label{PertPotential}
\end{equation}
Note that the first term provides no transition between quantum levels because it is position-independent, but clearly the matrix elements of the second term relates polymer wave functions with different quantum numbers, and can lead to transitions. In first order perturbation theory, the probability of the corresponding transition after an observation time $T$ is equal to
\begin{equation}
P _{n \rightarrow m} \left( T \right) = \frac{1}{\hbar ^{2}} \Big| \int _{0} ^{T} V _{nm} \left( t \right) dt \Big| ^{2} = \left( \frac{\mathcal{P} _{nm}}{\hbar } \right) ^{2} \Big| \int _{0} ^{T} \dot{h} e ^{-i \left( E _{n} - E _{m} \right) t / \hbar} dt \Big| ^{2} , \label{Pnm}
\end{equation}
with $\mathcal{P} _{nm} = i \frac{\hbar}{2 \lambda} \mathcal{T} _{nm}$, where the matrix elements
\begin{equation}
\mathcal{T} _{nm} = \sum _{\mu = 0} ^{\infty} \psi ^{(n)} _{\mu} \left( \widehat{U} _{\lambda} - \widehat{U} _{- \lambda} \right) \psi ^{(m)} _{\mu} \label{Tnm}
\end{equation}
encodes the transition between states. In \cref{MatrixElements} we explicitly evaluate these matrix elements in a simple fashion. The result is
\begin{equation}
2 \left( \varepsilon _{n} - \varepsilon _{m} \right) \mathcal{T} _{nm} = \frac{1}{2} \psi ^{(n)} _{1} \psi ^{(m)} _{1} - \upsilon ^{-2} \sum _{\mu = 0} ^{\infty} \mu \psi ^{(n)} _{\mu} \psi ^{(m)} _{\mu} , \label{Tnm2}
\end{equation}
where $\varepsilon _{n}$ are the polymer energy levels reported in \cref{table} and $\upsilon$ is defined in \cref{Dimensionless}. Now we focus on the quantum regime. In the limiting case $\lambda \ll l _{0}$ this equation becomes
\begin{equation}
\mathcal{P} _{nm} \approx i \frac{mg}{\omega _{nm}} \left[ 1 + \left( - 1 \right) ^{n-m} \frac{g}{2 l _{0}\omega _{nm} ^{2}}  \left( \frac{\lambda}{l _{0}} \right) ^{3} \right] , \label{Pnm2}
\end{equation}
where $\omega _{nm} = \left( E _{n} - E _{m} \right) / \hbar $ is the quantum-mechanical angular transition frequency. On the other hand, the integral in (\ref{Pnm}) makes no polymer corrections to the transition probabilities, but it depends on the level  of vibration noise in the spectrometer. Therefore, the lowest order polymer correction to the probability of transition is proportional to $s ^{-3}$. In the GRANIT experiment an accelerometer which is sensitive in the frequency range from $0$ to $500 \mbox{Hz}$, has been used to study the transition between the six lowest gravitational neutron quantum states \cite{Codau}. Unfortunately, this implies that the correction term in \cref{Pnm2} would be strongly suppresed in the GRANIT experiment. However, it could in principle be used to establish an alternative bound for $\lambda$. For comparison with experimental results, we write the probability of transition per unit time in terms of the vertical acceleration of the mirror
\begin{equation}
p ^{a} _{n \rightarrow m} \simeq \left( \frac{mg}{\hbar} \right) ^{2} \frac{1}{\omega ^{4} _{nm}} \left[ 1 + \left( - 1 \right) ^{n-m} \frac{g}{ l _{0}\omega _{nm} ^{2}}  \left( \frac{\lambda}{l _{0}} \right) ^{3} \right] S _{a} \left( \omega _{nm} \right) , \label{pnm}
\end{equation}
where
\begin{equation}
S _{a} \left( \omega \right) = \lim _{T \rightarrow \infty} \frac{1}{T} \Big| \int _{0} ^{T} \ddot{h} e ^{i \omega t} dt \Big| ^{2} \label{S}
\end{equation}
is the acceleration power spectrum \cite{Codau}. The lifetime for the $n$-th level can be calculated, summing the contributions (\ref{pnm}) for all transitions, namely $ \tau _{n} ^{-1} = \sum _{m \neq n} p ^{a} _{n \rightarrow m}$. In the problem at hand, the polymer quantum levels lifetime are given by
\begin{equation}
\tau _{n} = \frac{t _{n}}{1 + t _{n} \Omega _{n} \upsilon ^{-3}} , \label{lifetime}
\end{equation}
where $t _{n}$ is the expected lifetime of the $n$-th neutron gravitational quantum state due to vibrations of the mirror ($t _{n} \sim 10 ^{4} \mbox{s}$) \cite{Codau} and
\begin{equation}
\Omega _{n} = \left( \frac{mg}{\hbar} \right) ^{2} \frac{g}{l _{0}} \sum _{m \neq n} \left( - 1 \right) ^{n-m} \frac{S _{a} \left( \omega _{nm} \right)}{ \omega ^{6} _{nm}} . \label{Omega}
\end{equation}
As in the previous section, we can use this equation to establish an upper bound for the fundamental length as a function of the best precision measurements, i.e. $\Delta t _{n} < \Delta t ^{\mbox{exp}} _{n}$ with $\Delta t _{n} = \tau _{n} - t _{n}$. We obtain
\begin{equation}
\lambda ^{3} < l _{0} ^{3} \frac{\Delta t ^{\mbox{exp}} _{n}}{t _{n} ^{2} \vert \Omega _{n} \vert} , \label{BoundLambda2}
\end{equation}
where $\Delta t ^{\mbox{exp}} _{n}$ is the maximal experimental error in the measurement of $t _{n}$. Now we perform a numerical analysis based on the data reported in \cite{Codau} for the neutron lifetimes of neutron gravitational quantum states due to vibrations of the mirrors. First we assume that the maximal error bar in the measurements of $t _{1} = 10 ^{5} \mbox{s}$ (the larger lifetime) is in the same order that for the $\beta$ decay time, i.e. $\Delta t ^{\mbox{exp}} _{n} \sim 1 \mbox{s}$. The best precision measurement of the $\beta$ decay time of the neutrons is about $\left( 878.5 \pm 0.7 _{stat} \pm 0.3 _{sys} \right) \mbox{s}$. In order to evaluate $\Omega _{1}$, we assign the average acceleration power spectrum $\overline{S} _{a} \left( \omega _{nm} \right)  \sim 10 ^{-10} \mbox{m} ^{2} \mbox{Hz} ^{3}$ to all possible transitions. A numerical evaluation of the sum appearing in \cref{Omega} gives $\Omega _{1} \sim 10 ^{-3} \mbox{Hz}$. With these values, \cref{BoundLambda2} yields again $\lambda < 10^{-6} \mbox{m}$. This upper bound is in the same order as that obtained in the previous section when confronted with the GRANIT experiment for thermal neutrons free falling in the gravitational field. With an improvement of the experimental precision in time measurements, a better upper bound can be achieved.

\section{Transition Rate of a Polymer Bouncer} \label{TransitionRate}

The observation of spontaneous decay of an excited state in the GRANIT experiment would be of interest, since it would be a Planck-scale physics effect. This decay rate is too low, as estimated in Ref.\cite{Pignol}. Since the spectrum of a polymer bouncer is slightly shifted respect to the standard energy spectrum (\ref{EnergyShift2}), we expect the rate of this decay will change as a trace of quantum gravitational effects via the short-distance (discretized) structure at high energies. Nevertheless, although the polymer correction to the decay rate is expected to be extremely low, it is of conceptual and theoretical interest to estimate it. In what follows we discuss this issue, and we present the detailed calculations in \cref{TRPolymerBouncer}.

The quantum-mechanical transition rate for a bouncer to make a transition $k \rightarrow n$, in the quadrupole approximation, is
\begin{equation}
\Gamma _{k \rightarrow n} = \frac{4}{15} \frac{\omega _{kn} ^{5}}{M ^{2} _{pl} c ^{4}} Q _{kn} ^{2} , \label{TransRate}
\end{equation}
where $M _{pl}$ is the Planck mass, $\omega _{kn}$ is the angular frequency of transition and $Q _{kn} = m \left< \psi _{k} \right| z ^{2} \left| \psi _{n} \right> $ is the quadrupole moment of transition \cite{Pignol}. In the polymer framework, the first $\lambda$-dependent contribution is due to the shift in energy, and the second one arises from the polymer quadrupole moment. For the two lowest quantum states, we find that the probability of spontaneous graviton emission (at lowest order in $\lambda$) is
\begin{equation}
\Gamma ^{\lambda} _{2 \rightarrow 1} \simeq \Gamma _{2 \rightarrow 1} \left( 1 + \frac{1}{2} \frac{\lambda ^{2}}{l _{0} ^{2}}\right) . \label{21TransRate}
\end{equation}
So, there is a theoretical difference in the probability of espontaneous graviton emission due to the spatial ``grainy" structure. Although this phenomenon is further away to be detected in the laboratory (even the quantum-mechanical result), it is interesting from the conceptual point of view because the spontaneous decay of an excited state would be a true Planck-scale physics effect.

\section{Conclusions} \label{Conclusions}

In the present work, we have investigated the problem of a particle subject to the Earth's gravitational field (described by Newton's law) but assuming that space is quantized, as suggested by various candidates to Quantum Theory of Gravity (such as LQG, String theory, and noncommutative geometries). In this scenario, deviations from the usual quantum effects are induced by the spatial discreteness, but not by a new short-range gravitational interaction as in Refs.\cite{Nesvizhevsky,Frank}. We have called this problem ``\textit{The Polymer Bouncer}", by obvious reasons. We solved the polymer Schr\"{o}dinger equation in an analytical fashion, and we evaluated numerically the corresponding energy levels. We showed that the polymer energy spectrum is in a close agreement with the obtained by using perturbation theory, supporting the hypothesis that $\lambda \ll l _{0}$ in the quantum domain.

The implications of the introduction of a nonzero fundamental length scale in quantum theory are quite profound. For example, there is a belief that if quantum gravity effects are taken into account, the Heisenberg uncertainty relations should be modified. In polymer quantization the problem is subtle because the notion of discreteness. Unlike GUP theories, in PQ there is no deformation of the algebra of the observables, rather the Hilbert space is such that the momentum operator is realized only indirectly through the translation operator. In  \cref{PolymerBouncer} we found that polymer quantum mechanics at short enough wave lengths can exhibit classical behaviour, resembling to those found, on a different setup, by Magueijo and Smolin in deformed special relativity. This finding is also in agreement with the 't Hooft proposal about the possibility of a deterministic quantum mechanics at Planck scale.

Experiments with thermal neutrons free falling in the gravitational field offer an opportunity to test observations with signatures of quantum gravitational effects. In this frame, we have established an upper bound for the fundamental length of the polymer theory by confronting our results with the best measurements performed in the GRANIT experiment, namely $\lambda \ll 0.6 \buildrel _{\circ} \over {\mathrm{A}}$. It is commonly believed that $\lambda$ is in the order of the Planck length $ \sim 10 ^{-35} \mbox{m}$ (no known data substantiates this conjecture), which is $10 ^{24}$ times smaller than the upper bound we established. Therefore our results are not yet feasible for the energy scales of present experiments, but they motivate for the search of other mechanisms that could provide a better bound for the fundamental length. In \cref{SmallVibrations} we studied the polymer corrections to the transitions induced by small vibrations, and we found an upper bound of the same order as that obtained in \cref{GRANIT}. In both cases, we established a relation for the upper bound as a function of the maximal experimental error performed in the GRANIT experiment. The improvement of the precision in such experimental methods will produce better bounds according to \cref{BoundLambda,BoundLambda2}. For the completion of this work, in \cref{TransitionRate} we briefly studied the polymer corrections to the quantum-mechanical transition rate for a bouncer to make a transition $k \rightarrow n$. This phenomenon, although is further away to be detected in the laboratory (even the quantum-mechanical result), it is interesting from the conceptual point of view because the spontaneous decay of an excited state would be a true Planck-scale physics effect.

\appendix

\section{The Quantum Bouncer} \label{QuantumBouncer}

In this section we recall briefly the problem of a particle of mass $m$ bouncing on a perfectly reflecting surface under the influence of gravity, that is, of a particle in the potential
\begin{eqnarray}
V \left( z \right) = \left\lbrace \begin{array}{c} \;\;\ mgz , \;\;\;\;\;\;\ z > 0 \\ \;\;\;\;\ \infty , \;\;\;\;\;\;\;\;\;\ z < 0
\end{array} \right. . \label{GravPotential}
\end{eqnarray}
where $g$ is the Earth's gravitational field strength near the surface. The wave function $\psi \left( z \right) $ of a quantum bouncer obeys the stationary Schr\"{o}dinger equation for the vertical motion along $z$ axis
\begin{equation}
- \frac{\hbar ^{2}}{2m} \frac{\partial ^{2} \psi }{\partial z ^{2}} + m g z \psi  = E \psi . \label{SchroEquation}
\end{equation}
The solution must obey the following boundary conditions, $\psi \left( z \right) $ must vanish asymptotically as $z \rightarrow \infty$, and $\psi \left( z = 0 \right) = 0$ because of the presence of the mirror at $z=0$. Making the change of variables $x = \frac{z}{l _{0}} - \frac{E}{m g l _{0}}$, where $l _{0} = \left( \frac{\hbar ^{2}}{2m ^{2}g} \right) ^{1/3}$ is the characteristic gravitational length, \cref{SchroEquation} can be written as
\begin{equation}
\frac{d ^{2} \psi }{dx ^{2}} = x \psi . \label{AiryDiffEq}
\end{equation}
The general solution to \cref{AiryDiffEq} can be written in terms of the Airy functions, $Ai \left( x \right) $ and $Bi \left( x \right) $. Since the function $Bi \left( x \right) $ goes to infinity as its arguments grows, it is not an acceptable solution for this problem, where $z$ is unbounded from above. Then, the solution is of the form $\psi \left( x \right) = N Ai \left( x \right) $, where $N$ is an appropriate normalization factor. The second boundary condition, $\psi \left( z = 0 \right) = N  Ai \left( - \frac{E }{ m g l _{0}} \right)$, establishes the quantized energy levels of the stationary states
\begin{equation}
E _{n} = - m g l _{0} a _{n} , \label{EnergySpectrum}
\end{equation}
where $a _{n}$ is the $n$-th zero of the Airy function. The sequence of zeros of the Airy function has no simple analytic expression, but fairly good approximations can be obtained. For negative arguments, the Airy function is related to the Bessel functions by
\begin{equation}
Ai \left( - y \right) = \frac{\sqrt{y}}{3} \left[ J _{1/3} \left( \xi \right) + J _{-1/3} \left( \xi \right) \right] ,  \label{Airy-Bessel}
\end{equation}
where $\xi = \frac{2}{3} y ^{3/2}$. For the zeros $Ai \left( a _{n} \right) = 0 $, one obtains the following approximate solution
\begin{equation}
a _{n} \approx - \left[ \frac{3 \pi}{2} \left( n - \frac{1}{4} \right) \right] ^{2/3} . \label{AiryZeros}
\end{equation}
It is accurate within $1\%$, even for $n=1$, and it is exact in the semiclassical limit $n \gg 1$. Substitution of (\ref{AiryZeros}) into (\ref{EnergySpectrum}) establishes a good approximation for the energy spectrum.

The wave function for the $n$-th (non-degenerate) state reads
\begin{equation}
\psi _{n} \left( z \right) = N _{n} Ai \left( a _{n} + \frac{z}{l _{0}} \right) \theta \left( z \right) , \label{WaveEquation1}
\end{equation}
where $\theta \left( z \right) $ is the Heaviside step function, and $N _{n}$ is determined from the normalization condition,
\begin{equation}
\int _{0} ^{\infty} \vert \psi _{n} \left( z \right) \vert ^{2} dz = N _{n} ^{2} l _{0} \int _{a _{n}} ^{\infty} Ai ^{2} \left( y \right) dy = N _{n} ^{2} l _{0} Ai ^{\prime 2} \left( a _{n} \right) = 1 . \label{Normalization}
\end{equation}

\section{Matrix Elements $\mathcal{T} _{mn}$} \label{MatrixElements}

To calculate the matrix element $\mathcal{T} _{mn}$ of \cref{Tnm}, we start with the polymer Schr\"{o}dinger equation (\ref{PolSchrEq}) expressed in terms of the translation operator,
\begin{equation}
2 \varepsilon _{n} \psi _{\mu} ^{(n)} = \left( 2 - \widehat{U} _{\lambda} - \widehat{U} _{- \lambda} \right) \psi _{\mu} ^{(n)} + \upsilon ^{-1} \mu \psi _{\mu} ^{(n)}  . \label{PolSchrod2}
\end{equation}
One can multiply (\ref{PolSchrod2}) by $\left( \widehat{U} _{\lambda} - \widehat{U} _{- \lambda} \right) \psi _{\mu} ^{(m)}$, multiply the polymer Schr\"{o}dinger equation for the wave function by $\left( \widehat{U} _{\lambda} - \widehat{U} _{- \lambda} \right) \psi _{\mu} ^{(n)}$ to sum them and to calculate the sum for $\mu \in \mathbb{Z} ^{+}$. The result is
\begin{eqnarray}
2 \sum _{\mu = 0 } ^{\infty} \left[ \varepsilon _{n} \psi _{\mu} ^{(n)} \left( \widehat{U} _{\lambda} - \widehat{U} _{- \lambda} \right) \psi _{\mu} ^{(m)} + \left( n \leftrightarrow m \right) \right] = \sum _{\mu = 0} ^{\infty} \Big[ \left( \widehat{U} _{\lambda} - \widehat{U} _{- \lambda} \right) \psi _{\mu} ^{(n)} \left( 2 - \widehat{U} _{\lambda} - \widehat{U} _{- \lambda} \right) \psi _{\mu} ^{(m)} + \left( n \leftrightarrow m \right)  \Big] + \label{T} \\ \upsilon ^{-1} \sum _{\mu = 0} ^{\infty} \mu \left[ \psi _{\mu} ^{(n)} \left( \widehat{U} _{\lambda} - \widehat{U} _{- \lambda} \right) \psi _{\mu} ^{(m)} + \left( n \leftrightarrow m \right) \right] \nonumber
\end{eqnarray}
The left hand side (LHS) can be computed by using the following simple result
\begin{equation}
\sum _{\mu = 0 } ^{\infty} \psi _{\mu} ^{(n)} \widehat{U} _{\pm \lambda} \psi _{\mu} ^{(m)} = \sum _{\mu = 0 } ^{\infty} \psi _{\mu} ^{(m)} \widehat{U} _{\mp \lambda} \psi _{\mu} ^{(n)} . \label{Identity1}
\end{equation}
So we obtain
\begin{equation}
LHS = 2 \left( \varepsilon _{n} - \varepsilon _{m} \right) \sum _{\mu = 0 } ^{\infty} \psi _{\mu} ^{(n)} \left( \widehat{U} _{\lambda} - \widehat{U} _{- \lambda} \right) \psi _{\mu} ^{(m)} = 2 \left( \varepsilon _{n} - \varepsilon _{m} \right) \mathcal{T} _{nm}, \label{LHS}
\end{equation}
where $\mathcal{T} _{nm}$ is the required matrix element (\ref{Tnm}). The first term RHS1 in the right hand side of eq.(\ref{T}) can be studied as follows. We know that to the lowest order in $\lambda$  we have $\widehat{U} _{ \lambda} - \widehat{U} _{- \lambda} \simeq - 2 \lambda \frac{\partial}{\partial z}$ and $2 - \widehat{U} _{ \lambda} - \widehat{U} _{- \lambda} \simeq - \lambda ^{2} \frac{\partial ^{2}}{\partial z ^{2}}$. Therefore the following approximation is valid
\begin{eqnarray}
RHS1 & = & \sum _{\mu = 0} ^{\infty} \left( \widehat{U} _{ \lambda} - \widehat{U} _{- \lambda} \right) \psi _{\mu} ^{(n)} \left( 2 - \widehat{U} _{ \lambda} - \widehat{U} _{- \lambda} \right) \psi _{\mu} ^{(m)} + \left( n \leftrightarrow m \right) , \label{RHS1} \\ & \simeq & - \frac{1}{4} \left( \widehat{U} _{ \lambda} - \widehat{U} _{- \lambda} \right) \left[ \left( \widehat{U} _{ \lambda} - \widehat{U} _{- \lambda} \right) \psi _{\mu} ^{(n)} \left( \widehat{U} _{ \lambda} - \widehat{U} _{- \lambda} \right) \psi _{\mu} ^{(m)} \right] , \nonumber \\ & = & - \frac{1}{2} \left( \widehat{U} _{ \lambda} - \widehat{U} _{- \lambda} \right) \psi _{0} ^{(n)} \left( \widehat{U} _{ \lambda} - \widehat{U} _{- \lambda} \right) \psi _{0} ^{(m)} . \nonumber
\end{eqnarray}
Note that in this case $RHS1 = - \frac{1}{2} \psi ^{(n)} _{1} \psi ^{(m)} _{1}$ because the boundary condition $\psi ^{(n)} _{0} = 0$. Nevertheless, in the limiting case $\lambda \ll l _{0} $ becomes $RHS1 = - 2 \lambda ^{3} \psi ^{\prime} _{n} \left( 0 \right) \psi ^{\prime} _{m} \left( 0 \right)$, where $\psi _{n}$ is the quantum mechanical wave function \cref{SchroEquation}. 

For computing the second term RHS2 in the right hand side of eq.(\ref{T}) we use the formula
\begin{equation}
\sum _{\mu = 0 } ^{\infty} \mu \psi _{\mu} ^{(n)} \widehat{U} _{\pm \lambda} \psi _{\mu} ^{(m)} = \sum _{\mu = 0 } ^{\infty} \psi _{\mu} ^{(m)} \left( \mu \mp 1 \right) \widehat{U} _{\mp \lambda} \psi _{\mu} ^{(n)} . \label{Identity2}
\end{equation}
The term RHS2 can be written as
\begin{equation}
RHS2 = - \upsilon ^{-1} \sum _{\mu = 0} ^{\infty} \psi ^{(n)} _{\mu}  \left( \widehat{U} _{ \lambda} + \widehat{U} _{- \lambda} \right) \psi ^{(m)} _{\mu} . \label{RHS2}
\end{equation}
After multiplying the Schr\"{o}dinger equation \cref{PolSchrod2} by the wave function $\psi ^{(m)} _{\mu}$ and performing the sum for $\mu \in \mathbb{Z} ^{+}$, one obtains
\begin{equation}
\sum _{\mu = 0} ^{\infty} \psi ^{(n)} _{\mu}  \left( \widehat{U} _{ \lambda} + \widehat{U} _{- \lambda} \right) \psi ^{(m)} _{\mu} = \upsilon ^{-1} \sum _{\mu = 0} ^{\infty} \mu \psi ^{(n)} _{\mu} \psi ^{(m)} _{\mu} , \label{RHS2,2}
\end{equation}
where the orthogonality of the polymer wave functions have been used. Then RHS2 becomes
\begin{equation}
RHS2 = - \upsilon ^{-2} \sum _{\mu = 0} ^{\infty} \mu \psi ^{(n)} _{\mu} \psi ^{(m)} _{\mu} . \label{RHS2,3}
\end{equation}
The substitution of (\ref{LHS}), (\ref{RHS1}) and (\ref{RHS2,3}) into (\ref{T}) establishes \cref{Tnm2}. 

In the quantum-mechanical regime $\lambda \ll l _{0}$, we approximate the sum appearing in \cref{Tnm2} by using the Euler-Maclaurin formula,
\begin{equation}
\sum _{\mu = 0} ^{\infty} \mu \psi ^{(n)} _{\mu} \psi ^{(m)} _{\mu} \simeq \frac{1}{l _{0}} \int _{0} ^{\infty} z \psi _{n} \left( z \right) \psi _{m} \left( z \right) dz = \frac{2 \left( - 1 \right) ^{n-m}}{ \left( a _{n} - a _{m} \right) ^{2}} = 2 \left( - 1 \right) ^{n-m} \left( \frac{mg l _{0}}{\hbar \omega _{nm}} \right) ^{2} , \label{znm}
\end{equation}
where $a _{n}$ is the $n$-the zero of the Airy function and $\omega _{nm} = \left( E _{n} - E _{m} \right) / \hbar $ is the quantum-mechanical angular transition frequency. As discussed in this section \cref{LHS} becomes $RHS1 = - 2 \lambda ^{3} \psi ^{\prime} _{n} \left( 0 \right) \psi ^{\prime} _{m} \left( 0 \right)$ for $\lambda \ll l _{0}$. These results establish the required matrix elements $\mathcal{P} _{nm}$ (\cref{Pnm2}) at quantum level.

\section{Transition Rate of a Polymer Bouncer} \label{TRPolymerBouncer}

In this section we present the calculations of the transition rate of a Polymer Bouncer. The polymer transition frequency (in the first order approximation), $\omega ^{\lambda} _{kn}$ , is given by
\begin{equation}
\omega ^{\lambda} _{kn} = \omega _{kn} \left( 1 - \frac{a _{k} + a _{n}}{60 s ^{2}} \right) . \label{FrequencyTrans}
\end{equation}
This formula is obtained with the help of \cref{EnergyShift2}. On the other hand, we must evaluate the polymer quadrupole moment $Q ^{\lambda} _{kn} = m \lambda ^{2} \sum _{\mu} \mu ^{2} \psi ^{(k)} _{\mu} \psi ^{(n)} _{\mu}$. To this end we use first order perturbation theory. The perturbation shifts the wave function to
\begin{equation}
\left| \Psi _{n} \left( \lambda \right) \right> = \left| \psi _{n} \right> - \frac{\lambda ^{2}}{24 m \hbar ^{2}} \sum _{k \neq n} \frac{\left< \psi _{k} \right| p ^{4} \left| \psi _{n} \right>}{E _{n} - E _{k}} \left| \psi _{k}  \right> , \label{ShiftsWavFunc}
\end{equation}
where $p$ is the standard quantum-mechanical momentum operator. By using the Hamiltonian, $H = \frac{p ^{2}}{2m} + mgz $, the first order shift in the wave function becomes
\begin{equation}
\left| \Psi _{n} \left( \lambda \right) \right> = \left| \psi _{n} \right> - \left( \frac{\lambda}{l _{0}} \right) ^{3} \sum _{l \neq n} \left( -1 \right) ^{l-n} F _{ln} \left| \psi _{l}  \right> , \label{ShiftsWavFunc2}
\end{equation}
where we have defined
\begin{equation}
F _{ln} = \frac{1}{3 \left( a _{k} - a _{n} \right) ^{3}} \left[ a _{n} - \frac{6}{ \left( a _{k} - a _{n} \right) ^{2}} \right] . \label{Fln}
\end{equation}
For the evaluation of the expectation value appearing in \cref{ShiftsWavFunc} we have used the well know results
\begin{equation}
\left< \psi _{k} \right| z \left| \psi _{n} \right> = \frac{2 \left( -1 \right) ^{n-k}}{\left( a _{k} - a _{n} \right) ^{2}} l _{0} \;\;\;\;\ , \;\;\;\;\ \left< \psi _{k} \right| z ^{2} \left| \psi _{n} \right> = \frac{24 \left( -1 \right) ^{k-n-1}}{\left( a _{k} - a _{n} \right) ^{4}} l _{0} ^{2} . \label{ZZ2}
\end{equation}
With the help of \cref{ShiftsWavFunc2}, the polymer quadrupole moment can be written in terms of the quantum-mechanical quandrupole moment ($Q _{kn} =  m \left< \psi _{k} \right| z ^{2} \left| \psi _{n} \right> $) as follow
\begin{equation}
Q ^{\lambda} _{kn} \simeq Q _{kn} - \left( \frac{\lambda}{l _{0}} \right) ^{2} \left[ \sum _{l \neq k} \left( -1  \right) ^{l-k} F _{lk} Q _{kn} + \sum _{l \neq n} \left( -1  \right) ^{l-n} F _{ln} Q _{kl} \right] . \label{PolQuadrupole}
\end{equation}
The substitution of \cref{FrequencyTrans,PolQuadrupole} into \cref{TransRate} produces a general expression for the transition rate of a polymer bouncer. For the transition between the two lowest quantum states, $2 \rightarrow 1$, we can evaluate numerically the sums involved in \cref{PolQuadrupole}. The approximate final result is \cref{21TransRate}.

\section*{Acknowledgements}
A. Mart\'{i}n-Ruiz would like to thank Valery Nesvizhevsky for the warm hospitality during his visit to the \textit{Institute Laue-Langevin}. AF acknowledges support from DGAPA-UNAM and CONACYT. LFU and AMR acknowledge support from the project DGAPA-UNAM-IN109013.

\end{document}